\documentclass[a4paper,11pt]{article}
\pdfoutput=1 

\usepackage{jcappub} 

\usepackage[T1]{fontenc} 
\usepackage{graphicx}

\title{\boldmath Constraining $\mu$ and $y$ distortions in the Cosmic Microwave Background with COBE/FIRAS Data}

\author[a]{Somita Dhal,}
\author[b]{Koustav Konar }
\author[a,1]{and R. K. Paul \note{Corresponding author.}}

\affiliation[a]{Birla Institute of Technology, Department of Physics, \\ 835215 Mesra, India.}
\affiliation[b]{Ruhr University Bochum, Faculty of Physics and Astronomy, Astronomical Institute (AIRUB), German Centre for Cosmological Lensing, \\ 44780 Bochum, Germany}

\emailAdd{somitadhal1995@gmail.com}
\emailAdd{koustav.konar@ruhr-uni-bochum.de}
\emailAdd{ratan\_bit1@rediffmail.com}

\abstract{This paper presents a novel approach to constrain the $\mu$- and y- distortions in the Cosmic Microwave Background (CMB) using the COBE/FIRAS data. The analysis draws from the concept of blackbody radiation inversion (BRI), a mathematical technique typically used to determine the temperature distribution from a radiated power spectrum. We study the deviations from the ideal blackbody spectrum or the spectral distortions by incorporating first a non-zero chemical potential $\mu$ via the Bose-Einstein distribution and then the Compton parameter $y$ while keeping the monopole temperature constant. We infer the results as probability distribution functions on these distortions. Finally, we derive $\mu = (8.913 \pm 0.736) \times 10^{-5}$ and $y = (1.532 \pm 0.092) \times 10^{-5}$ at a $68\%$ confidence interval. The results are consistent with prior values and provide tighter constraints on the CMB spectral distortion and synergies of the primordial Universe.}

\begin{document}
\maketitle
\flushbottom

\section{Introduction}

The cosmic microwave background (CMB) spectrum as measured by the Far Infrared Absolute Spectrophotometer on board the Cosmic Background Explorer satellite (COBE/FIRAS) is remarkably close to a perfect blackbody spectrum, a theoretical prediction based on the assumption that the CMB is the leftover radiation from a very hot, dense Universe that has cooled down over time, thus providing conclusive evidence for the Big Bang model. Subsequent studies, however, focused on the deviation from this ideal behaviour. These deviations, called spectral distortions, have been instrumental in probing the physics of the era of recombination \cite{Sunyaev_Zeldo_1969, Sunyaev_distortion_1970, Burigana_evolution_distortion_1991, Wayne_thermalisation_1993, chluba_evolution_distortion_2012, Lucca_synergy_2019}. Physical processes such as interactions of photons with matter, energy injection or exotic physics beyond the standard cosmological model are considered to be the primary producers \cite{Chluba_2014, fu_unlocking_2021}. Hence, understanding these distortions is crucial for a comprehensive knowledge of the thermal history and has implications for the cosmological model. 

The deviations from the ideal behaviour are parameterised and one of the key parameters used to quantify these distortions in the CMB spectrum is the chemical potential, denoted by $\mu$ \cite{chluba_teasing_2014}. As the efficiency of the photon non-conservation processes like double Compton and bremsstrahlung decays with cosmic expansion, photon energy density changes whereas the photon number density remains unchanged. This effectively introduces a non-zero chemical potential characterised by the $\mu$-distortion \cite{khatri_beyond_2012, bianchini_distortion_2022}. Secondly, the Compton y-parameter, denoted by $y$, quantifies the inverse Compton scattering of CMB photons off free electrons, primarily occurring within hot ionised regions such as galaxy clusters. The y-parameter is closely related to the thermal energy of the scattering electrons and can be used to probe the properties of cosmic structures and astrophysical processes \cite{Sunyaev_Zeldo_1969, Tashiro_CMB_spectra_distortion_2014}. Hence, CMB and its spectral distortion present a unique window into the Universe's early evolution and the underlying physics governing its dynamics. We can test cosmological models, constrain fundamental parameters and probe physics beyond the standard model by comparing theoretical predictions with observational data from experiments such as those with the COBE satellite. Prior studies of $\mu$- and $y$-distortions with the COBE/FIRAS data have put an upper limit. However, more precise constraints can reveal a better budget for the energy transfer in the early Universe as spectral distortions are sensitive to the synergies. Hence, renewed interests have emerged with the next generation of CMB surveys like BISOU \cite{BISOU_maffei2023} and concepts such as PIXIE \cite{kogut_PIXIE_2011, Kogut_PIXIE_goal_2016}, APSERa \cite{rao_APSERa_2015} and COSMO \cite{masi_COSMO_2023} with over a magnitude improvement on the sensitivity compared to COBE. Recent studies have shown the potential implications of such precise measurements \cite{chluba_distortion_2019, chluba_new_horizons_2021}. 

While awaiting the first data release, we re-examine the COBE data to estimate the $\mu$- and $y$-distortions with an inversion technique. We substitute a probability distribution function (PDF) for the distortion in the Bose-Einstein distribution and infer the results via Blackbody Radiation Inversion (BRI) \cite{Bojarski_BRI_1982}. This involves solving a Fredholm integral, which is executed by a change of variable. The procedure of BRI and its application on the CMB spectra has been presented in \cite{Koustav_revisiting_2021}. Furthermore, BRI has been used to calculate the temperature of the CMB monopole spectrum in the context of non-extensive thermostatics \cite{dhal2024study} and to constrain the mass of photon \cite{bhattacharjee2024utilizing}. Here, we follow the same while considering how the Plack law reformulates under the $\mu$- and $y$-distortions.

The manuscript is structured as follows: Section \ref{sec:procedure} introduces the basics of $\mu$- and $y$-distortions and lays out the general analysis pipeline used in the calculation. In Section \ref{sec:results}, we present the major results.  Finally, in Section \ref{sec:discussion}, we summarise the findings with a discussion.

\section{Procedure}
\label{sec:procedure}

\subsection{Chemical potential and \texorpdfstring{$\mu$}{mu}-distortion}
\label{subsec:mu}

Blackbody radiation refers to the electromagnetic radiation emitted by an idealised object known as a blackbody. Planck's law is a mathematical formula that describes the energy density of the radiation emitted per unit wavelength at a given temperature while the blackbody is in thermal equilibrium. Crucially, it assumes a vanishing chemical potential ($\mu=0$) following non-conservation of photon number and minimisation of Gibbs free energy. However, photons do not interact with each other and require interaction with other particles to maintain equilibrium which affects the argument of chemical potential being zero \cite{wurfel_chemical_potential_1982}. This scenario appears at a redshift range of $5 \times 10^4 \lesssim z \lesssim 2 \times10^6$ where bremsstrahlung and double Compton scattering become inefficient and the energy injection results in a non-zero chemical potential \cite{Burigana_evolution_distortion_1991, Sunyaev_small_scale_1970}. Under this circumstance, an ideal blackbody spectrum could not be formed and therefore, we choose a Bose-Einstein (BE) distribution as a static solution for Compton scattering with a dimensionless non-zero chemical potential ($\mu$)
\begin{align}
\label{eq:be_law}
    f_\mathrm{BE} = \frac{1}{e^{r+\mu} - 1} \;, 
\end{align}
where $r=h\nu/kT$ and $\nu$ is the frequency, $h$ is the Planck's constant, $c$ is the speed of light, $k$ is the Boltzmann's constant and $T$ is the temperature of the CMB. We use $T = 2.72548 \pm 0.00057$ K \cite{fixen_cmb_temp_2009}. The intensity of radiation can be written as
\begin{align}
\label{eq:planck_law}
    I(\nu, \mu) = \frac{2h\nu^3}{c^2} \frac{1}{e^{r+\mu} - 1} \;.
\end{align}
For brevity, we rearrange the prefactor and define
\begin{equation}
\label{eq:g_definition}
    G(\nu, \mu) := \frac{c^2}{2h\nu^3} I(\nu, \mu)\;.
\end{equation}
Then, Equation \ref{eq:planck_law} yields
\begin{align}
\label{eq:planck_redefined}
    G(\nu, \mu) = \frac{1}{e^{r+\mu} - 1} \;.
\end{align}
Now, for a probabilistic approach of inferring $\mu$, we introduce a PDF for $\mu$, denoted by $\alpha(\mu)$, and marginalise Equation \ref{eq:planck_redefined} by integrating over all possible values of $\mu$ within a broad enough prior range of $\mu \in [\mu_1, \mu_2]$.
\begin{align}
\label{eq:g_integral_1}
    G(\nu, \mu) = \int_{\mu_1}^{\mu_2} \frac{1}{e^{r+\mu} - 1} \alpha(\mu) \;\mathrm{d}\mu \;.
\end{align}
The goal is to calculate $\alpha(\mu)$. However, Equation \ref{eq:g_integral_1} is a Fredholm integral equation known to be an ill-posed problem, i.e. the output is very sensitive to the fluctuations in the input. Historically, multiple methods have been employed to solve such integrals including among others reparameterisation and entropy maximisation \cite{tikhonov_1977_solutions, lakhtakia_inverse_1986, Chen_mobius_1990, dou_maximum_1992, JiPing_2006_black, cheng_variational_2008, jieer_GMRES_2013}.  

A recent study has proposed a new approach that overcomes this limitation by assuming a probability distribution with unknown parameters \cite{Koustav_revisiting_2021}. This approach allows for an effective and robust inversion of the spectral data using only a few data points. Here, we follow the same prescription and solve the Fredholm equation by employing a change of variable
\begin{align}
\label{eq:M_definition}
    \mu = \mu_1 + (\mu_2 - \mu_1) M \;.
\end{align} 
The newly introduced variable $M$ falls within the range of values from 0 to 1. We substitute $M$ in Equation \ref{eq:g_integral_1}
\begin{align}
\label{eq:g_integral_2}
    G(\nu, \mu) = \int_{\mu_1}^{\mu_2} \frac{1}{e^{r+[\mu_1 + (\mu_2 - \mu_1) M]} - 1} \alpha([\mu_1 + (\mu_2 - \mu_1) M]) \;\mathrm{d}M \;.
\end{align}
We now define a Gaussian function for the variable $M$ 
\begin{align}
\label{eq:pdf_mu_definiton}
    w(M) := \alpha([\mu_1 + (\mu_2 - \mu_1) M]) = m \cdot \mathrm{exp}\left[-\left( \frac{M-n}{p}\right)^2 \right] \;,
\end{align}
where $m$ denotes the overall amplitude, $n$ denotes the peak position of the Gaussian function and $p$ denotes the full width at half maximum. Substituting $w(M)$ in Equation \ref{eq:g_integral_2} returns
\begin{align}
\label{eq:g_integral_3}
    G(\nu, M) = \int_0^1 \frac{\mu_2 - \mu_1}{e^{r+[\mu_1 + (\mu_2 - \mu_1) M]} - 1} w(M) \;\mathrm{d}M \;.
\end{align}
Now, we changed the $\mu$ dependence with $M$. However, the substitution still requires the integration limits on $\mu$. In principle, as long as the limits are broad enough to cover the whole prior on $\mu$, the integration in Equation \ref{eq:g_integral_2} should work. Here, we intend to calculate the $\alpha(\mu)$. The Full Width at a tenth of the Maximum (FWTM) for a Gaussian distribution with standard deviation $\sigma$ is $4.29193 \sigma$. We motivate the choice of the order from the previous value of $\mu = 9 \times 10^{-5}$ obtained using COBE data \cite{fixen_1996_cobe}. For the range, we choose a reasonably broad $\sigma = 0.8 \times 10^{-5}$. With that, the limits are
\begin{align}
    \mu_1 = \left( 9  - \frac{1}{2} \cdot 0.8 \cdot 4.29193 \right) \times 10^{-5} = 7.2 \times 10^{-5} \;, \\
    \mu_2 = \left( 9  + \frac{1}{2} \cdot 0.8 \cdot 4.29193 \right) \times 10^{-5} = 1 \times 10^{-4} \;.
\end{align}
Additionally, Equation \ref{eq:g_integral_1} decays rapidly with increasing frequency due to the denominator. The narrower limits are in place due to the negligible contribution from the PDF beyond these points. However, for clarification, we have checked the variation of results for different $\mu_1$ and $\mu_2$ in Section \ref{sec:results}.

Our objective is to determine the distribution $w(M)$ in Equation \ref{eq:g_integral_3}, which following the definition in Equation \ref{eq:M_definition} returns the PDF of $\mu$, which we explicitly write
\begin{align}
\label{eq:alpha_mu}
    \alpha(\mu) = m \cdot \mathrm{exp}\left[-\left( \frac{\frac{\mu - 7.2 \times 10^{-5}}{2.8 \times 10^{-5}} -n}{p}\right)^2 \right] \;.
\end{align}
The experimental value of $G(\nu)$ on the left-hand side of Equation \ref{eq:g_integral_3} can be calculated using
\begin{align}
\label{eq:g_calculated_from_monopole}
    G(\nu) = \frac{c^2}{2h\nu^3} I_\mathrm{M} \;,
\end{align}
where $I_\mathrm{M}$ is the value of the monopole intensity taken from the COBE/FIRAS dataset \footnote{\href{https://lambda.gsfc.nasa.gov/product/cobe/firas_monopole_get.html}{https://lambda.gsfc.nasa.gov/product/cobe/firas\_monopole\_get.html}} described in \cite{fixen_1996_cobe, fixen_mather_2002_cobe}. 

Similarly, we substitute Equation \ref{eq:pdf_mu_definiton} on the right-hand side of Equation \ref{eq:g_integral_3}. Then, a set of three equations is derived for a corresponding set of three frequencies. We now have three simultaneous integral equations with three unknowns. The solution results in the determination of values for $m, n,$ and $p$ in Equation \ref{eq:pdf_mu_definiton}, which are then transformed to $\mu$. The same process is repeated for another nine sets of frequencies and the calculated $m, n,$ and $p$ values are plugged into Equation \ref{eq:alpha_mu}. All the corresponding values are shown in Table \ref{table:mu_distortion_data}.  In total, we have ten PDFs of $\mu$, which we denote via $a_i(\mu)$'s.

\begin{figure}
    \centering
    \includegraphics[width=0.75\textwidth]{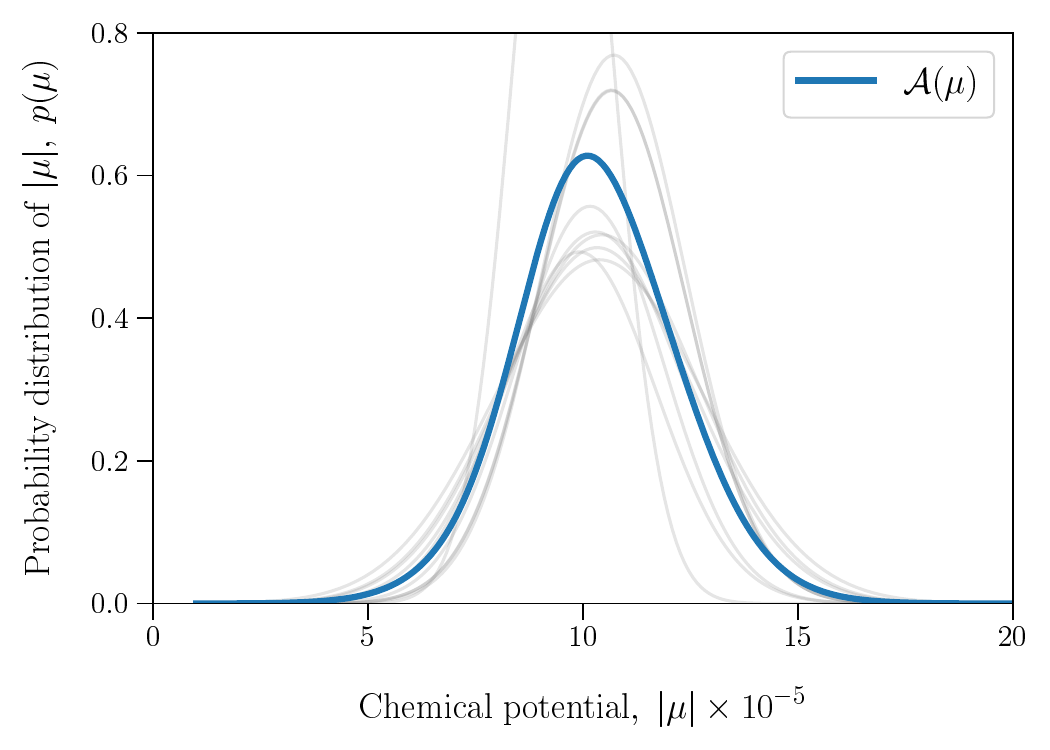}
    \caption{The PDFs of the $\mu$-distortion calculated from Table \ref{table:mu_distortion_data}. The grey lines are the 10 individual PDFs and the blue line is the average.}
    \label{fig:mu_distortion}
    \vspace{.3cm}
\end{figure}

\begin{table} [h!]
\centering
\renewcommand{\arraystretch}{1.3}
\begin{tabular}{c|c|c|c|c}
\hline
\textbf{Frequency set ($\times 10^{11}$ Hz)} & $\boldsymbol{m}$ & $\boldsymbol{n}$ & $\boldsymbol{p}$ & \textbf{PDFs} \\
\hline

0.681, 0.954, 1.224	& 0.482	& 1.136	& 1.222	& $a_1(\mu)$ \\
1.498, 1.770, 2.043	& 1.297	& 0.836	& 0.570	& $a_2(\mu)$ \\
2.724, 2.994, 3.267	& 0.521	& 1.103 & 1.022 & $a_3(\mu)$ \\
2.859, 3.132, 3.402	& 0.499	& 1.117	& 1.119	& $a_4(\mu)$ \\
3.267, 3.541, 3.813	& 0.719	& 1.237	& 0.857	& $a_5(\mu)$ \\
3.948, 4.221, 4.491	& 0.557	& 1.058 & 0.861	& $a_6(\mu)$ \\
4.083, 4.356, 4.629	& 0.720	& 1.235	& 0.858	& $a_7(\mu)$ \\
2.724, 3.132, 3.540	& 0.517	& 1.155	& 1.115	& $a_8(\mu)$ \\
4.083, 4.491, 4.902	& 0.493	& 0.968	& 0.902	& $a_9(\mu)$ \\
5.445, 5.583, 6.261	& 0.769	& 1.256	& 0.839	& $a_{10}( \mu)$ \\
\hline
\end{tabular}
\caption{The values of $m, n$ and $p$ for different probability functions of the $\mu$ distribution are listed. Each corresponds to the calculation from a distinct frequency set in the first column.}
\label{table:mu_distortion_data}
\end{table}
We then take the average of all the PDFs, i.e. $\sum_{i=1}^{10} a_i(\mu)/10$, which yields the final probability distribution for the $\mu$-distortion. We denote the average as $\mathcal{A}(\mu)$ and is shown in Figure \ref{fig:mu_distortion}. 

\subsection{Compton parameter and \texorpdfstring{$y$}{y}-distortion}
\label{subsec:y}
\begin{figure}  
    \centering
    \includegraphics[width=0.75\textwidth]{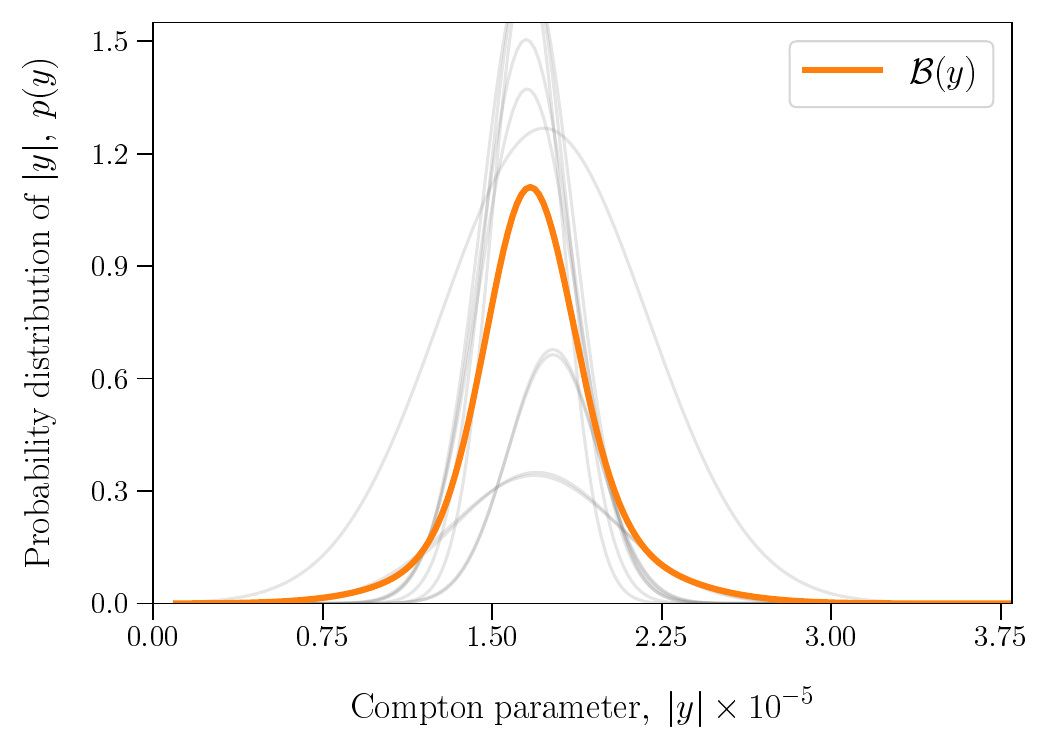}
    \caption{The PDF of the $y$-distortion calculated from Table \ref{table:y_distortion_data}. The grey lines are the 10 individual PDFs and the orange line is the average.}
    \label{fig:y_distortion}
    \vspace{.3cm}
\end{figure}

Compton scattering maintains the Bose-Einstein spectrum in the $\mu$-distortion era. However, its effectiveness is a function of photon energy and decreases with reducing photon energy. Hence, energy injections between $10^3 \lesssim z \lesssim 5 \times 10^4$ lead to the scattering of low-energy photons off the high-energetic electrons \cite{khatri_beyond_2012}. In this inverse Compton scattering process, the photons acquire energy leading to a shift of the CMB spectra to higher frequencies, which is defined via the $y$-distortion and known as the Sunyaev-Zel'dovich effect (typically $y \ll 1$) \cite{Sunyaev_small_scale_1970}. This effect continues at lower redshift values where the CMB photons interact with energetic electrons in galaxy clusters. The value of $y$ can be obtained using \cite{Biyajima_2012}
\begin{align}
\label{eq:compton_definition}
    y = \frac{4\pi (I_\mathrm{M} - I_\mathrm{BE}) (e^r -1)^2}{C_\mathrm{B} \nu^3 c r e^r \left(r \coth{(r/2)} - 4 \right)} \;,
\end{align}
where $I_\mathrm{M}$ is the monopole spectra measured by COBE, $I_\mathrm{BE}$ is the Bose-Einstein spectra defined via Equation \ref{eq:planck_law} and $C_\mathrm{B} = 8\pi h/c^3$. After rearranging, we write
\begin{align}
    I_\mathrm{M}(\nu, y) = \frac{2h \nu^3}{c^2} \left[ \frac{1}{e^{r+\mu} - 1} + \frac{C_\mathrm{B} \nu^3 c r e^r \left(r \coth{(r/2)} - 4 \right) y}{\left(e^r - 1\right)^2} \right] \;.
\end{align}
Now applying blackbody radiation inversion to the above equation, we get
\begin{align}
\label{eq:y_dist}
    I_\mathrm{M}(\nu, y) =  \int_{y_1}^{y_2} \frac{2h \nu^3}{c^2} \left[  \frac{1}{e^{r+\mu} - 1} + \frac{ r e^r \left(r \coth{(r/2)} - 4 \right) y}{\left(e^r - 1\right)^2} \right] \beta(y)\;\mathrm{d}y \;,
\end{align}
where $\beta(y)$ is the probability distribution of $y$. We follow the same procedure as Section \ref{subsec:mu} but with a $y$ dependence, introduce $G(\nu, y) = c^2 I_\mathrm{M}/2h \nu^3$ and change of variables $y= y_1 + (y_2 -y_1) Y$. Then, Equation \ref{eq:y_dist} is written as
\begin{align}
    G(\nu, y) = \int_0^1 \left[\frac{1}{e^{r+\mu} - 1} + \frac{ r e^r \left(r \coth{(r/2)} - 4 \right) y}{\left(e^r - 1\right)^2} \right] (y_2 -y_1) \beta(y_1 + (y_2 - y_1)Y) \; \mathrm{d}Y \;.
\end{align}
Similarly, following the previous upper limit of $|y| < 1.5 \times 10^{-5}$ from the literature \cite{fixen_1996_cobe} and $\sigma = 0.8 \times 10^{-5}$, we set the integration limits as $y_1 = 1.328 \times 10^{-5}$ and $y_2 = 1.672 \times 10^{-5}$. The PDF of $y$-distortion is now
\begin{align}
\label{eq:alpha_y}
    \beta(y) = \Tilde{m} \cdot \mathrm{exp}\left[ -\left( \frac{\left(\frac{y - 1.328 \times 10^{-5}}{3.440 \times 10^{-6}} \right) - \Tilde{n}}{\Tilde{p}}\right)^2 \right] \;.
\end{align}
The same procedure as in Section \ref{subsec:mu} with $y$ dependence yields 10 PDFs, denoted by $b_i(y)$'s and we again take the average. All the values of $\Tilde{m}, \Tilde{n}$ and $\Tilde{p}$ are given in Table \ref{table:y_distortion_data}. The final distribution, $\mathcal{B}(y) = \sum_{i=1}^{10} b_i(y)/10$, is shown in Figure \ref{fig:y_distortion}.

\begin{table} [h!]
\centering
\renewcommand{\arraystretch}{1.3}
\begin{tabular}{c|c|c|c|c}
\hline
\textbf{Frequency set ($\times 10^{11}$ Hz)} & $\boldsymbol{\Tilde{m}}$ & $\boldsymbol{\Tilde{n}}$ & $\boldsymbol{\Tilde{p}}$ & \textbf{PDFs} \\
\hline
0.681, 0.954, 1.22	& 1.505	& 0.938 & 0.853	& $b_1(\mu)$ \\
1.498, 1.770, 2.04	& 1.373	& 0.950 & 0.885 & $b_2(\mu)$ \\
2.724, 2.994, 3.267	& 0.664	& 1.284 & 0.830	& $b_3(\mu)$ \\
2.859, 3.132, 3.402	& 0.350	& 1.072 & 1.471	& $b_4(\mu)$ \\
3.267, 3.541, 3.813	& 1.720	& 0.948 & 0.737	& $b_5(\mu)$ \\
3.948, 4.221, 4.491	& 1.708	& 0.948 & 0.838	& $b_6(\mu)$ \\
4.083, 4.356, 4.629	& 0.342	& 1.046 & 1.532	& $b_7(\mu)$ \\
2.724, 3.132, 3.540	& 0.678	& 1.283 & 0.816	& $b_8(\mu)$ \\
4.083, 4.491, 4.902	& 1.701	& 0.948 & 0.638	& $b_9(\mu)$ \\
5.445, 5.583, 6.261	& 1.268	& 1.167 & 1.884 & $b_{10}( \mu)$ \\
\hline
\end{tabular}
\caption{The values of $\Tilde{m}, \Tilde{n}$ and $\Tilde{p}$ for different PDFs of the $y$-distortion are listed. Each corresponds to the calculation from a distinct frequency set in the first column.}
\label{table:y_distortion_data}
\end{table}

\section{Results}
\label{sec:results}
In the previous section, we calculated the PDFs on $\mu$- and $y$-distortions using BRI. In this section, we present the principal findings of our analysis from those PDFs. We start with the $\mu$-distortion and then describe the $y$-distortion. 

\subsection{\texorpdfstring{$\mu$}{mu}-distortion}
The effect of non-zero chemical potential on the CMB spectra is characterised by $\mu$-distortion. We have utilised the BE distribution to describe the distorted spectra and implemented the BRI method to determine a probability distribution on $\mu$. The resulting PDF is denoted by $\mathcal{A}(\mu)$ and is the average of the ten PDFs that we calculated from ten different frequency sets. The normalised $\mathcal{A}(\mu)$ is $\alpha(\mu)$ in Equation \ref{eq:pdf_mu_definiton} we set out to calculate. In that regard, we normalise $\mathcal{A}(\mu)$ via
\begin{align}
    \alpha(\mu) = \frac{\mathcal{A}(\mu)}{\int_{7.2 \times 10^{-5}}^{1 \times 10^{-4}} \mathcal{A}(\mu) \;\mathrm{d}\mu} = 0.878\times10^{5} \cdot \mathcal{A}(\mu) \;. 
\end{align}
The first moment or the mean is calculated as 
\begin{align}
    \overline{\mu} = \int_{7.2 \times 10^{-5}}^{1 \times 10^{-4}} \mu \cdot \alpha(\mu) \;\mathrm{d}\mu = 8.913 \times 10^{-5} \;.
\end{align}
The second moment, i.e. the variance, is
\begin{align}
    \sigma^2_\mu = \int_{7.2 \times 10^{-5}}^{1 \times 10^{-4}} (\mu - \overline{\mu})^2 \cdot \alpha(\mu) \;\mathrm{d}\mu = 5.424 \times 10^{-11} \;.
\end{align}
From which the standard deviation is written as
\begin{align}
    \Delta\mu = \sigma_\mu = \sqrt{\sigma^2_\mu} = 7.365 \times 10^{-6} \;.
\end{align}
This is the best constraint from our analysis, $\mu = (8.913 \pm 0.736)\times 10^{-5}$, which is fully consistent with the previous $\mu < 9 \times 10^{-5}$ limit set in \cite{fixen_1996_cobe}. 

We can extend these one-point statistics to higher-order moments. The third ordered moment is called \textit{skewness} and quantifies the degree of asymmetry in the distribution of data points;
\begin{align}
    \gamma_{1, \mu} = \frac{1}{\sigma^3_\mu} \int_{7.2 \times 10^{-5}}^{1 \times 10^{-4}} (\mu - \overline{\mu})^3 \cdot \alpha(\mu) \;\mathrm{d}\mu = - 0.461 \;.
\end{align}
In a normal distribution, the skewness is expected to be 0. Negative skewness refers to a statistical scenario where the distribution is skewed to the left. This implies that the tail of the distribution extends further to the left of the peak (or mean) compared to the right side \cite{groeneveld_skewness_1984, DeCarlo_kurtosis_1997}. This suggests the mean value of $\mu$ is tending towards lower values, further indicating that the PDF is consistent with the previous upper limit and small value of $\mu$.

The fourth moment measures the heaviness of the tails and the peakedness of the distribution relative to a normal distribution and is called \textit{kurtosis}.
\begin{align}
    \gamma_{2, \mu} = \frac{1}{\sigma^4_\mu} \int_{7.2 \times 10^{-5}}^{1 \times 10^{-5}} (\mu - \overline{\mu})^4 \cdot \alpha(\mu) \;\mathrm{d}\mu = 2.486 \;.
\end{align}
The kurtosis of a distribution relative to that of a normal distribution, i.e. the excess kurtosis, is typically evaluated. The kurtosis of a normal distribution is 3, so excess kurtosis is calculated by subtracting 3 from the actual kurtosis value. Our result yields a positive number which implies that the peak of the curve is slightly higher than the normal \cite{DeCarlo_kurtosis_1997}.

\begin{figure}
    \centering
    \includegraphics[width=0.75\textwidth]{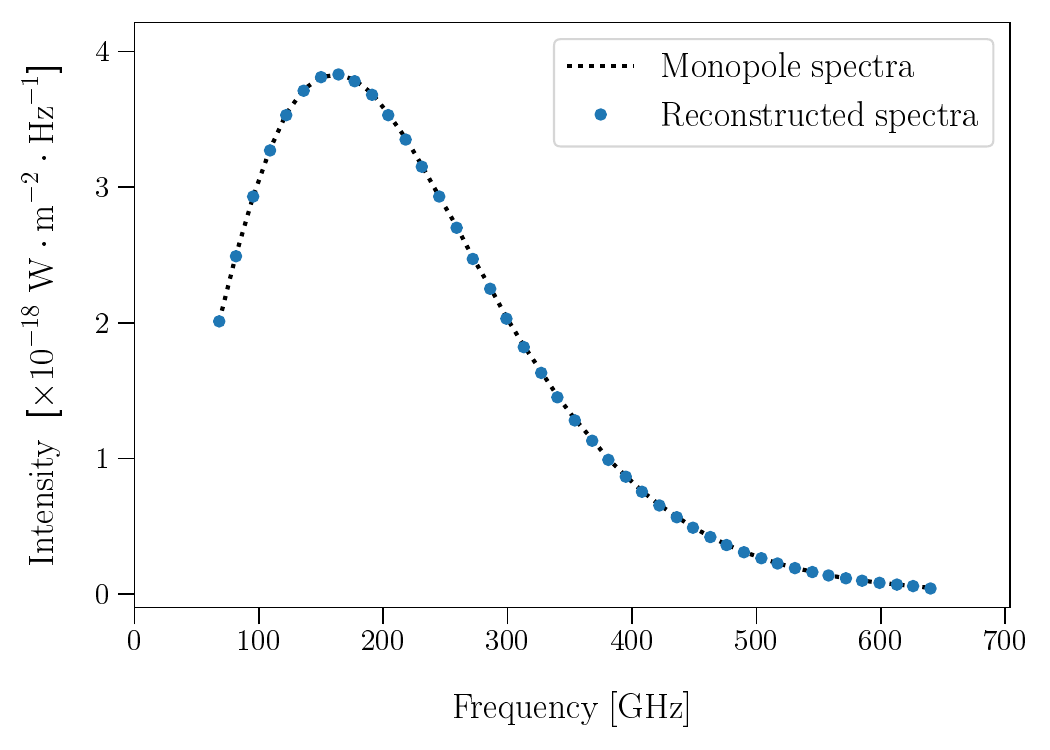}
    \caption{The figure shows the experimental and reconstructed intensities plotted against frequency for the calculated value of $\mu$. There is a significant similarity between the reconstructed and COBE/FIRAS dataset.}
    \label{fig:mu_reconstruction}
    \vspace{.3cm}
\end{figure}

For validation, we reconstructed the radiation intensity by utilising the calculated $\alpha(\mu)$ in Equation \ref{eq:g_integral_1}. Figure \ref{fig:mu_reconstruction} displays a comparison between the original spectrum data from COBE/FIRAS and the reconstructed data. We note that the errors on the data, i.e. $\Delta \mathrm{I}/\mathrm{I} \lesssim 10^{-5}$ and hence not shown in the figure.

We quantify the goodness-of-fit with a $\chi^2$ measurement via
\begin{align}
\label{eq:chi_square}
    \chi^2 = \sum w_i \left[\mathrm{R}_i - \mathrm{O}_i \right]^2 \;,
\end{align}
where $w_i$ are the weights, $R_i$ are the reconstructed data and $O_i$ are the original data points. The weights are determined by $1/\sigma_i^2$ on each data point. The reduced chi-square $(\chi^2/\mathrm{dof})$, with $\mathrm{dof}$ representing the number of degrees of freedom, is calculated to be 1.04, which implies an excellent fit to the original data. The full dataset with input and reconstructed spectra is provided in Table \ref{table:reconstructed_data}.

In Section \ref{subsec:mu}, we set narrow limits on the integration ($\mu_1$ and $\mu_2$) considering the pre-existing value of $\mu$ and the decaying nature of the denominator in Equation \ref{eq:g_integral_1}. Now, we change the limits and observe the effects on the constraint of $\mu$. Additionally, we incorporate experimental errors in intensities $\Delta \mathrm{I}/\mathrm{I}$ associated with these selected frequencies in Equation \ref{eq:g_calculated_from_monopole}. The values are listed in Table \ref{table:mu_variation_data}. We observe that the error bars overlap and varying the integral limits does not change our constraints significantly, i.e. our results are stable. 

\begin{table} [h!]
\centering
\renewcommand{\arraystretch}{1.8}
\begin{tabular}{c|c|c}
\multicolumn{3}{c}{\textbf{Variation in result by changing the integral range}} \\
\hline
$\boldsymbol{\mu_1 (\times 10^{-5})}$ & $\boldsymbol{\mu_2 (\times 10^{-5})}$ & $\boldsymbol{\mathrm{Calculated \;\mu} (\times 10^{-5})}$ \\
\hline
7.5   & 10.5   & $8.856 \pm 0.663$ \\  
8   & 11   & $9.176 \pm 0.807$ \\ 
6 & 10 & $8.346 \pm 1.095$ \\
6.5   & 11.5   & $9.953 \pm 1.219$ \\
\hline
\multicolumn{3}{c}{\textbf{Variation in result by incorporating $\Delta \mathrm{I}$}} \\
\hline
\multicolumn{3}{c}{$\mathrm{Calculated \;\mu} = (8.913 \pm 0.736) \times 10^{-5} $}

\end{tabular}
\caption{Variation in the calculation of $\mu$-distortion with changing values of the integration limit $\mu_1, \mu_2$ and incorporating the error of intensity $\Delta\mathrm{I}$.}
\label{table:mu_variation_data}
\end{table}

\subsection{\texorpdfstring{$y$}{y}-distortion}
\begin{figure}
    \centering
    \includegraphics[width=0.75\textwidth]{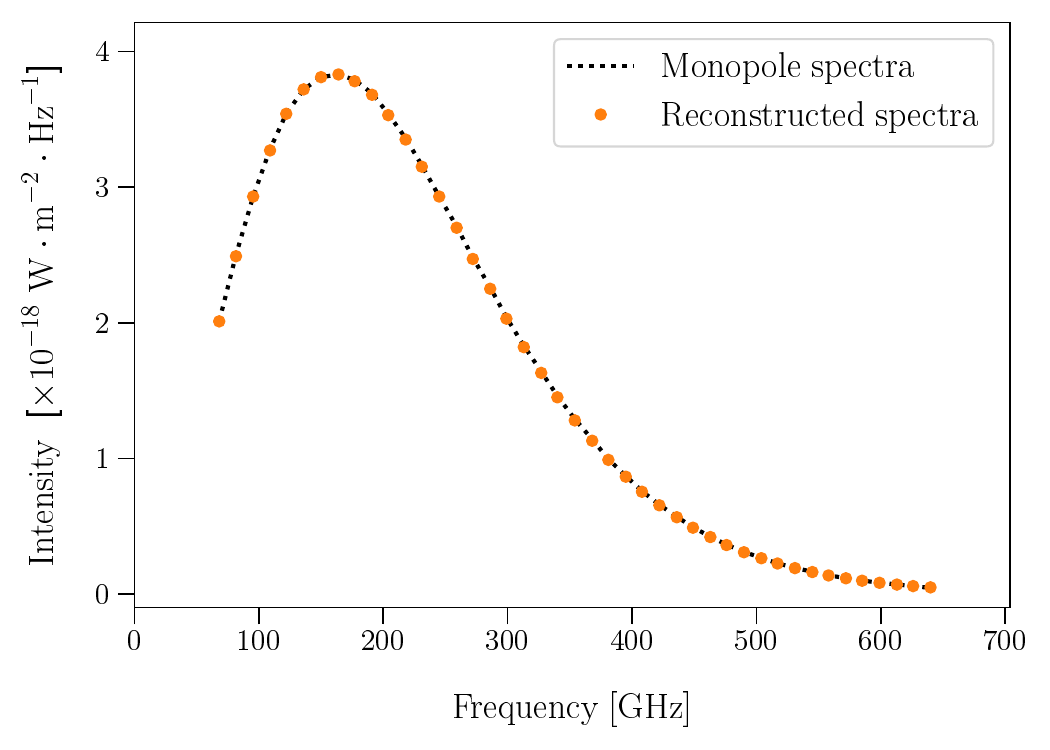}
    \caption{The figure illustrates the experimental and reconstructed intensities graphed against frequency using the inferred value of $y$. It demonstrates how accurate the reconstructed data are by comparing them with the COBE/FIRAS dataset.}
    \label{fig:y_reconstruction}
    \vspace{.3cm}
\end{figure}

The Compton $y$ parameter describes how the CMB spectrum shifts to higher frequencies following the scattering of low-energetic photons with high-energetic electrons in galaxy clusters. We modelled this distorted spectrum as a deviation from the Bose-Einstein distribution described in \cite{Biyajima_2012} and applied the BRI methods which resulted in the PDF. We start by normalising the average PDF, $\mathcal{B}(y)$, we get from Section \ref{subsec:y}.
\begin{align}
    \beta(y) = \frac{\mathcal{B}(y)}{\int_{1.326 \times 10^{-5}}^{1.671 \times 10^{-5}} \mathcal{B}(y) \;\mathrm{d}y} = 3.766 \times 10^5 \cdot \mathcal{B}(y) \;. 
\end{align}
The mean is evaluated as
\begin{align}
    \overline{y} = \int_{1.326 \times 10^{-6}}^{1.671 \times 10^{-5}} y \cdot \beta(y) \;\mathrm{d}y = 1.532 \times 10^{-5} \;. 
\end{align}
The variance, similarly, yields
\begin{align}
    \sigma^2_y = \int_{1.326 \times 10^{-6}}^{1.671 \times 10^{-5}} (y - \overline{y})^2 \cdot \beta(y) \;\mathrm{d}y = 8.482 \times 10^{-13} \;. 
\end{align}
The standard deviation of $y$ is then
\begin{align}
    \Delta y = \sigma_y = \sqrt{\sigma^2_y} = 9.210 \times 10^{-7} \;.
\end{align}
Hence, the $y$-distortion is calculated as $y=(1.532 \pm 0.092) \times 10^{-5}$, which is fully consistent with prior value of $y=1.5 \times 10^{-5}$ in \cite{fixen_1996_cobe, Dhal_calculation_2023, Dhal_dipole_2023} using the same dataset. Likewise, the input spectra can be reconstructed. The result is shown in Figure \ref{fig:y_reconstruction} and all the values are presented in Table \ref{table:reconstructed_data}. The input as well as the reconstructed spectra only deviate with $\Delta \mathrm{I}/\mathrm{I} \lesssim 10^{-5}$. The $\chi^2/\mathrm{dof}$ in this case is 1.05, which corresponds to a good fit.

Next, in Table \ref{table:y_variation_data}, we present the variation in the estimation of the $y$-distortion when the integration limits are changed and the spectral uncertainties are accounted for. All the values are consistent with each other and the integration limits have negligible effect.

\begin{table} [h!]
\centering
\renewcommand{\arraystretch}{1.8}
\begin{tabular}{c|c|c}
\multicolumn{3}{c}{\textbf{Variation in result by changing the integral range}} \\
\hline
$\boldsymbol{y_1 (\times 10^{-5})}$ & $\boldsymbol{y_2 (\times 10^{-5})}$ & $\boldsymbol{\mathrm{Calculated \;} y (\times 10^{-5})}$ \\
\hline
1.3   & 1.7   & $1.582 \pm 0.099$ \\  
1.2  & 1.6   & $1.595 \pm 0.095$ \\ 
1.25 & 1.75 & $1.529 \pm 0.090$ \\
1.1  & 1.8   & $1.686 \pm 0.123$ \\
\hline
\multicolumn{3}{c}{\textbf{Variation in result by incorporating $\Delta \mathrm{I}$}} \\
\hline
\multicolumn{3}{c}{$\mathrm{Calculated \;} y = (1.532 \pm 0.092) \times 10^{-5} $}

\end{tabular}
\caption{Variation in the calculation of $y$-distortion with changing values of the integration limit $y_1, y_2$ and incorporating the error of intensity $\Delta\mathrm{I}$.}
\label{table:y_variation_data}
\end{table}

\section{Discussion}
\label{sec:discussion}
The cosmic microwave background (CMB) radiation is the remnant of the hot Big Bang model radiated at a redshift of $z \sim 1100$ following the decoupling of matter and radiation. By definition, it makes CMB the earliest direct signal we have to study the primordial universe. Early studies focus on the nature and profile of the average monopole intensities. The data release of COBE/FIRAS allowed for comparison with an ideal blackbody and the deviation owing to the sensitivity of the instruments. These differences are called spectral distortions. Spectral distortion signals are rich in information about the energy transfer at a high redshift and have been studied extensively to constrain cosmology and particle physics models \cite{hu_thermalisation_1993, chluba_evolution_distortion_2012, shaw_primordial_magnetism_2012, Lucca_synergy_2019, fu_unlocking_2021} 

In this paper, we study two distortions, namely $\mu$-distortion specifying the deviation due to non-zero chemical potential and $y$-distortion characterising the interaction between photons and energetic electrons. We modelled the respective spectra using the Bose-Einstein distribution and its deviation due to inverse Compton scattering. The models are then implemented within the blackbody radiation inversion (BRI) framework by which we calculated probability distribution functions for $\mu$- and $y$-distortions using the COBE/FIRAS dataset \cite{fixen_1996_cobe, fixen_mather_2002_cobe}.

Different moments of the PDFs are calculated which is presented as our principle findings as $\mu = (8.913 \pm 0.736) \times 10^{-5}$ and $y=(1.532 \pm 0.092) \times 10^{-5}$ at a $68\%$ confidence interval. The results are in excellent agreement with the prior limits obtained from the same dataset \cite{fixen_1996_cobe, Dhal_calculation_2023, Dhal_dipole_2023}. Higher-order moments revealed that the PDFs are slightly different than a standard normal distribution. The probability density function has a higher mass in the direction of the prior literature values, thus, reinforcing our results. Additionally, we investigated the effect of the integration limits on the inferred results and found them to be stable. Finally, we validated our analysis by reconstructing the input spectrum with the calculated value of $\mu$- and $y$-distortions. The goodness-of-fit was measured via the reduced $\chi^2$-test and the reconstructions have $ \chi^2/\mathrm{dof} =1.04$ for $\mu$- and $ \chi^2/\mathrm{dof} =1.05$ for $y$-distortions. This implies that the reconstructed data are an excellent fit to the input data. 

With the next generation of CMB projects such as BISOU \cite{BISOU_maffei2023}, PIXIE \cite{kogut_PIXIE_2011, Kogut_PIXIE_goal_2016}, APSERa \cite{rao_APSERa_2015} and COSMO \cite{masi_COSMO_2023} along with Stage-IV surveys like Simons Observatory \cite{ade_simons_obs_2019} and LiteBIRD \cite{paoletti_LiteBIRD_2022} on the horizon, studies are expected to observe better constraints on the energy transfers in the early Universe. We showed that the BRI is applicable and results are in agreement with prior values. Moreover, as our analysis requires a magnitude less data and is computationally inexpensive, it presents a possibility to analyse the spectral distortion with a stable and fast inversion technique. We have used the sky-averaged COBE data and there is a significant contamination in the $\mu$-distortion signal emanating from the lower redshift $y$-distortion signal. This problem has been studied where the authors found the constraint change based on the sky fraction used as input \cite{khatri_constraint_2015}. As a next step, it would be interesting to evaluate how the BRI constraints change with a more precise dataset and sky maps.

\appendix
\section{Appendix}
All the data products from this work are presented here in Table \ref{table:reconstructed_data}. The inputs are the first three columns, i.e. the frequency and the monopole intensity along with the error. The reconstructed data are calculated using the $\mu$- and $y$-distortion values. We also provide the respective errors associated with this process.

\begin{table} 
\centering
\resizebox{\textwidth}{!}{
\begin{tabular}{c|c|c|c|c|c|c}
\hline
\textbf{Frequency [GHz]} & $\boldsymbol{\mathrm{I}_\mathrm{mono}} \; [\times 10^{-18}]$ & $\boldsymbol{\mathrm{\Delta I}_\mathrm{mono}}\; [\times 10^{-23}]$ & $\boldsymbol{\mathrm{I}_\mu} \; [\times 10^{-18}]$ & $\boldsymbol{\mathrm{\Delta I}_\mu}\; [\times 10^{-23}]$ & $\boldsymbol{\mathrm{I}_y} \; [\times 10^{-18}]$ & $\boldsymbol{\mathrm{\Delta I}_y}\; [\times 10^{-23}]$ \\

\hline

$68.10$ & $2.01$ & $14.00$ & $2.01$ & $4.17$ & $2.01$ &$1.11$\\
$81.60$ & $2.50$ & $19.00$ & $2.49$ & $4.74$ & $2.49$&$1.44$ \\
$95.40$ & $2.93$ & $25.00$ & $2.93$ & $5.22$ & $2.93$ &$1.72$\\
$109.00$ & $3.28$ & $23.00$ & $3.27$ & $5.57$ & $3.27$&$1.92$ \\
$122.00$ & $3.54$ & $22.00$ & $3.53$ & $5.80$ & $3.54$ &$2.02$\\
$136.00$ & $3.72$ & $21.00$ & $3.71$ & $5.93$ & $3.72$ &$2.02$\\
$150.00$ & $3.81$ & $18.00$ & $3.81$ & $5.95$ & $3.81$ &$1.90$\\
$164.00$ & $3.83$ & $18.00$ & $3.83$ & $5.88$ & $3.83$ &$1.67$\\
$177.00$ & $3.79$ & $16.00$ & $3.78$ & $5.74$ & $3.78$ &$1.34$\\
$191.00$ & $3.69$ & $14.00$ & $3.68$ & $5.53$ & $3.68$ &$0.94$\\
$204.00$ & $3.54$ & $13.00$ & $3.53$ & $5.28$ & $3.53$ &$0.47$\\
$218.00$ & $3.36$ & $12.00$ & $3.35$ & $4.97$ & $3.35$ &$0.01$\\
$231.00$ & $3.16$ & $11.00$ & $3.15$ & $4.66$ & $3.15$ &$0.51$\\
$245.00$ & $2.93$ & $10.00$ & $2.93$ & $4.31$ & $2.93$ &$1.02$\\
$259.00$ & $2.71$ & $11.00$ & $2.70$ & $3.96$ & $2.70$ &$1.49$\\
$272.00$ & $2.48$ & $12.00$ & $2.47$ & $3.63$ & $2.47$ &$1.93$\\
$286.00$ & $2.26$ & $14.00$ & $2.25$ & $3.29$ & $2.25$ &$2.31$\\
$299.00$ & $2.04$ & $16.00$ & $2.03$ & $2.98$ & $2.03$ &$2.63$\\
$313.00$ & $1.83$ & $18.00$ & $1.82$ & $2.67$ & $1.82$ &$2.90$\\
$327.00$ & $1.64$ & $22.00$ & $1.63$ & $2.37$ & $1.63$ &$3.10$\\
$340.00$ & $1.46$ & $22.00$ & $1.45$ & $2.12$ & $1.45$ &$3.24$\\
$354.00$ & $1.29$ & $23.00$ & $1.28$ & $1.87$ & $1.28$ &$3.32$\\
$368.00$ & $1.14$ & $23.00$ & $1.13$ & $1.64$ & $1.13$ &$3.36$ \\
$381.00$ & $0.99$ & $23.00$ & $0.99$ & $1.44$ & $0.99$ &$3.34$\\
$395.00$ & $0.87$ & $22.00$ & $0.86$ & $1.26$ & $0.86$ &$3.28$\\
$408.00$ & $0.76$ & $21.00$ & $0.75$ & $1.10$ & $0.75$ &$3.20$\\
$422.00$ & $0.66$ & $20.00$ & $0.65$ & $0.95$ & $0.65$ &$3.08$\\
$436.00$ & $0.57$ & $19.00$ & $0.57$ & $0.82$ & $0.57$ &$2.94$\\
$449.00$ & $0.49$ & $19.00$ & $0.49$ & $0.71$ & $0.49$ &$2.78$\\
$463.00$ & $0.42$ & $19.00$ & $0.42$ & $0.61$ & $0.42$ &$2.62$\\
$476.00$ & $0.36$ & $21.00$ & $0.36$ & $0.53$ & $0.36$ &$2.44$\\
$490.00$ & $0.31$ & $23.00$ & $0.31$ & $0.45$ & $0.31$ &$2.27$\\
$504.00$ & $0.27$ & $26.00$ & $0.26$ & $0.38$ & $0.26$ &$2.09$\\
$517.00$ & $0.23$ & $28.00$ & $0.22$ & $0.33$ & $0.22$ &$1.92$\\
$531.00$ & $0.19$ & $30.00$ & $0.19$ & $0.28$ & $0.19$ &$1.76$\\
$545.00$ & $0.16$ & $32.00$ & $0.16$ & $0.23$ & $0.16$ &$1.6$\\
$558.00$ & $0.14$ & $33.00$ & $0.14$ & $0.20$ & $0.14$ &$1.44$\\
$572.00$ & $0.12$ & $35.00$ & $0.12$ & $0.17$ & $0.12$ &$1.30$\\
$585.00$ & $0.10$ & $41.00$ & $0.10$ & $0.14$ & $0.10$ &$1.17$\\
$599.00$ & $0.08$ & $55.00$ & $0.08$ & $0.12$ & $0.08$ &$1.05$\\
$613.00$ & $0.07$ & $88.00$ & $0.07$ & $0.10$ & $0.07$ &$0.93$\\
$626.00$ & $0.06$ & $155.00$ & $0.06$ & $0.09$ & $0.06$ &$0.83$\\
$640.00$ & $0.05$ & $282.00$ & $0.04$ & $0.07$ & $0.05$ &$0.73$\\
\hline
\end{tabular}
}
\caption{The full COBE dataset is presented in the first two columns where the frequencies are in \textbf{GHz}. All the intensities are converted into SI unit of $\boldsymbol{\mathrm{W} \cdot \mathrm{m}^{-2} \cdot \mathrm{Hz}^{-1}}$. The columns denoted by $\sigma$ correspond to the $68\%$ error in the spectra of the monopole and reconstruction using $\mu$- and $y$-distortion values from Section \ref{sec:results} respectively.}
\label{table:reconstructed_data}
\end{table}

\newpage
\acknowledgments

The authors would like to thank the Department of Physics at Birla Institute of Technology, Mesra, Ranchi, for offering an excellent research environment. Our heartfelt appreciation also goes to B. Pathak from the Department of Physics, BIT Mesra, Ranchi, for his constant encouragement in research. S. Dhal thanks UGC (Savitribai Jyotirao Phule Single Girl Child with grant number - UGCES-22-GE-ORI-F-SJSGC-3962) for the fellowship to carry out research.



\bibliographystyle{JHEP}
\bibliography{References.bib}

\providecommand{\href}[2]{#2}\begingroup\raggedright\begin{thebibliography}{10}

\bibitem{Sunyaev_Zeldo_1969}
Y.B.~{Zeldovich} and R.A.~{Sunyaev}, \emph{{The Interaction of Matter and Radiation in a Hot-Model Universe}}, \href{https://doi.org/10.1007/BF00661821}{\emph{apss} {\bfseries 4} (1969) 301}.

\bibitem{Sunyaev_distortion_1970}
R.A.~{Sunyaev} and Y.B.~{Zeldovich}, \emph{{The Spectrum of Primordial Radiation, its Distortions and their Significance}}, {\emph{Comments on Astrophysics and Space Physics} {\bfseries 2} (1970) 66}.

\bibitem{Burigana_evolution_distortion_1991}
C.~{Burigana}, L.~{Danese} and G.~{de Zotti}, \emph{{Formation and evolution of early distortions of the microwave background spectrum - A numerical study}}, {\emph{aap} {\bfseries 246} (1991) 49}.

\bibitem{Wayne_thermalisation_1993}
W.~Hu and J.~Silk, \emph{Thermalization and spectral distortions of the cosmic background radiation}, \href{https://doi.org/10.1103/PhysRevD.48.485}{\emph{Phys. Rev. D} {\bfseries 48} (1993) 485}.

\bibitem{chluba_evolution_distortion_2012}
J.~{Chluba} and R.A.~{Sunyaev}, \emph{{The evolution of CMB spectral distortions in the early Universe}}, \href{https://doi.org/10.1111/j.1365-2966.2011.19786.x}{\emph{mnras} {\bfseries 419} (2012) 1294} [\href{https://arxiv.org/abs/1109.6552}{{\ttfamily 1109.6552}}].

\bibitem{Lucca_synergy_2019}
M.~Lucca, N.~Sch\"oneberg, D.C.~Hooper, J.~Lesgourgues and J.~Chluba, \emph{{The synergy between CMB spectral distortions and anisotropies}}, \href{https://doi.org/10.1088/1475-7516/2020/02/026}{\emph{JCAP} {\bfseries 02} (2020) 026} [\href{https://arxiv.org/abs/1910.04619}{{\ttfamily 1910.04619}}].

\bibitem{Chluba_2014}
J.~Chluba, \emph{{Science with CMB spectral distortions}},  in \emph{{49th Rencontres de Moriond on Cosmology}}, pp.~327--334, 2014 [\href{https://arxiv.org/abs/1405.6938}{{\ttfamily 1405.6938}}].

\bibitem{fu_unlocking_2021}
H.~Fu, M.~Lucca, S.~Galli, E.S.~Battistelli, D.C.~Hooper, J.~Lesgourgues et~al., \emph{Unlocking the synergy between cmb spectral distortions and anisotropies}, {\emph{Journal of Cosmology and Astroparticle Physics} {\bfseries 2021} (2021) 050}.

\bibitem{chluba_teasing_2014}
J.~Chluba and D.~Jeong, \emph{Teasing bits of information out of the cmb energy spectrum}, \href{https://doi.org/10.1093/mnras/stt2327}{\emph{Monthly Notices of the Royal Astronomical Society} {\bfseries 438} (2014) 2065}.

\bibitem{khatri_beyond_2012}
R.~Khatri and R.A.~Sunyaev, \emph{Beyond y and $\mu$: the shape of the cmb spectral distortions in the intermediate epoch, $1.5\times 10\textsuperscript{4}\lesssim z\lesssim 2\times 10\textsuperscript{5}$}, {\emph{Journal of Cosmology and Astroparticle Physics} {\bfseries 2012} (2012) 016}.

\bibitem{bianchini_distortion_2022}
F.~Bianchini and G.~Fabbian, \emph{Cmb spectral distortions revisited: A new take on $\ensuremath{\mu}$ distortions and primordial non-gaussianities from firas data}, \href{https://doi.org/10.1103/PhysRevD.106.063527}{\emph{Phys. Rev. D} {\bfseries 106} (2022) 063527}.

\bibitem{Tashiro_CMB_spectra_distortion_2014}
H.~{Tashiro}, \emph{{CMB spectral distortions and energy release in the early universe}}, \href{https://doi.org/10.1093/ptep/ptu066}{\emph{Progress of Theoretical and Experimental Physics} {\bfseries 2014} (2014) 06B107}.

\bibitem{BISOU_maffei2023}
B.~Maffei, M.~Abitbol, N.~Aghanim, J.~Aumont, E.~Battistelli, J.~Chluba et~al., \emph{Bisou: a balloon project to measure the cmb spectral distortions},  in \emph{The Sixteenth Marcel Grossmann Meeting on Recent Developments in Theoretical and Experimental General Relativity, Astrophysics and Relativistic Field Theories: Proceedings of the MG16 Meeting on General Relativity; 5--10 July 2021}, pp.~1633--1644, World Scientific, 2023.

\bibitem{kogut_PIXIE_2011}
A.~Kogut, D.~Fixsen, D.~Chuss, J.~Dotson, E.~Dwek, M.~Halpern et~al., \emph{The primordial inflation explorer (pixie): a nulling polarimeter for cosmic microwave background observations}, {\emph{Journal of Cosmology and Astroparticle Physics} {\bfseries 2011} (2011) 025}.

\bibitem{Kogut_PIXIE_goal_2016}
A.~{Kogut}, J.~{Chluba}, D.J.~{Fixsen}, S.~{Meyer} and D.~{Spergel}, \emph{{The Primordial Inflation Explorer (PIXIE)}},  in \emph{Space Telescopes and Instrumentation 2016: Optical, Infrared, and Millimeter Wave}, H.A.~{MacEwen}, G.G.~{Fazio}, M.~{Lystrup}, N.~{Batalha}, N.~{Siegler} and E.C.~{Tong}, eds., vol.~9904 of \emph{Society of Photo-Optical Instrumentation Engineers (SPIE) Conference Series}, p.~99040W, July, 2016, \href{https://doi.org/10.1117/12.2231090}{DOI}.

\bibitem{rao_APSERa_2015}
M.~{Sathyanarayana Rao}, R.~{Subrahmanyan}, N.~{Udaya Shankar} and J.~{Chluba}, \emph{{On the Detection of Spectral Ripples from the Recombination Epoch}}, \href{https://doi.org/10.1088/0004-637X/810/1/3}{\emph{apj} {\bfseries 810} (2015) 3} [\href{https://arxiv.org/abs/1501.07191}{{\ttfamily 1501.07191}}].

\bibitem{masi_COSMO_2023}
S.~{Masi}, E.~{Battistelli}, P.~{de Bernardis}, F.~{Columbro}, A.~{Coppolecchia}, G.~{D'Alessandro} et~al., \emph{{The COSmic Monopole Observer (COSMO)}},  in \emph{The Sixteenth Marcel Grossmann Meeting. On Recent Developments in Theoretical and Experimental General Relativity, Astrophysics, and Relativistic Field Theories}, R.~{Ruffino} and G.~{Vereshchagin}, eds., pp.~1654--1671, July, 2023, \href{https://doi.org/10.1142/9789811269776_0131}{DOI}.

\bibitem{chluba_distortion_2019}
J.~{Chluba}, A.~{Kogut} and e.~{Patil}, \emph{{Spectral Distortions of the CMB as a Probe of Inflation, Recombination, Structure Formation and Particle Physics}}, \href{https://doi.org/10.48550/arXiv.1903.04218}{\emph{baas} {\bfseries 51} (2019) 184} [\href{https://arxiv.org/abs/1903.04218}{{\ttfamily 1903.04218}}].

\bibitem{chluba_new_horizons_2021}
J.~{Chluba}, M.H.~{Abitbol}, N.~{Aghanim} and e.~{Ali-Ha{\"\i}moud}, Y., \emph{{New horizons in cosmology with spectral distortions of the cosmic microwave background}}, \href{https://doi.org/10.1007/s10686-021-09729-5}{\emph{Experimental Astronomy} {\bfseries 51} (2021) 1515} [\href{https://arxiv.org/abs/1909.01593}{{\ttfamily 1909.01593}}].

\bibitem{Bojarski_BRI_1982}
N.~Bojarski, \emph{Inverse black body radiation}, \href{https://doi.org/10.1109/TAP.1982.1142844}{\emph{IEEE Transactions on Antennas and Propagation} {\bfseries 30} (1982) 778}.

\bibitem{Koustav_revisiting_2021}
K.~{Konar}, K.~{Bose} and R.K.~{Paul}, \emph{{Revisiting cosmic microwave background radiation using blackbody radiation inversion}}, \href{https://doi.org/10.1038/s41598-020-80195-3}{\emph{Scientific Reports} {\bfseries 11} (2021) 1008}.

\bibitem{dhal2024study}
S.~Dhal and R.~Paul, \emph{A study of cosmic microwave background using non-extensive statistics}, {\emph{Experimental Astronomy} {\bfseries 57} (2024) 25}.

\bibitem{bhattacharjee2024utilizing}
P.~Bhattacharjee, S.~Dhal and R.~Paul, \emph{Utilizing blackbody radiation inversion to attain an upper bound on the mass of photon using cosmic microwave background radiation}, {\emph{The European Physical Journal Plus} {\bfseries 139} (2024) 1}.

\bibitem{wurfel_chemical_potential_1982}
P.~{Wurfel}, \emph{{The chemical potential of radiation}}, \href{https://doi.org/10.1088/0022-3719/15/18/012}{\emph{Journal of Physics C Solid State Physics} {\bfseries 15} (1982) 3967}.

\bibitem{Sunyaev_small_scale_1970}
R.A.~{Sunyaev} and Y.B.~{Zeldovich}, \emph{{Small-Scale Fluctuations of Relic Radiation}}, \href{https://doi.org/10.1007/BF00653471}{\emph{apss} {\bfseries 7} (1970) 3}.

\bibitem{fixen_cmb_temp_2009}
D.J.~{Fixsen}, \emph{{The Temperature of the Cosmic Microwave Background}}, \href{https://doi.org/10.1088/0004-637X/707/2/916}{\emph{apj} {\bfseries 707} (2009) 916} [\href{https://arxiv.org/abs/0911.1955}{{\ttfamily 0911.1955}}].

\bibitem{tikhonov_1977_solutions}
A.N.~Tikhonov and V.Y.~Arsenin, \emph{Solutions of ill-posed problems}, V. H. Winston \& Sons, Washington, D.C.: John Wiley \& Sons, New York (1977).

\bibitem{lakhtakia_inverse_1986}
M.N.~{Lakhtakia} and A.~{Lakhtakia}, \emph{{On some relations for the inverse blackbody radiation problem}}, \href{https://doi.org/10.1007/BF00697419}{\emph{Applied Physics B: Lasers and Optics} {\bfseries 39} (1986) 191}.

\bibitem{Chen_mobius_1990}
N.-x.~Chen, \emph{Modified m\"obius inverse formula and its applications in physics}, \href{https://doi.org/10.1103/PhysRevLett.64.1193}{\emph{Phys. Rev. Lett.} {\bfseries 64} (1990) 1193}.

\bibitem{dou_maximum_1992}
L.~Dou and R.~Hodgson, \emph{Maximum entropy method in inverse black body radiation problem}, {\emph{Journal of applied physics} {\bfseries 71} (1992) 3159}.

\bibitem{JiPing_2006_black}
J.~Ye, F.~Ji, T.~Wen, X.-X.~Dai, J.-X.~Dai and W.E.~Evenson, \emph{The black-body radiation inversion problem, its instability and a new universal function set method}, {\emph{Physics Letters A} {\bfseries 348} (2006) 141}.

\bibitem{cheng_variational_2008}
J.~Cheng and T.~Zhou, \emph{A variational expectation-maximization method for the inverse black body radiation problem}, {\emph{Journal of Computational Mathematics} (2008) 876}.

\bibitem{jieer_GMRES_2013}
J.~Wu and Z.~Ma, \emph{A regularized gmres method for inverse blackbody radiation problem}, \href{https://doi.org/10.2298/TSCI110316078W}{\emph{Thermal Science} {\bfseries 17} (2013) 847}.

\bibitem{fixen_1996_cobe}
D.J.~{Fixsen}, E.S.~{Cheng}, J.M.~{Gales}, J.C.~{Mather}, R.A.~{Shafer} and E.L.~{Wright}, \emph{{The Cosmic Microwave Background Spectrum from the Full COBE FIRAS Data Set}}, \href{https://doi.org/10.1086/178173}{\emph{apj} {\bfseries 473} (1996) 576} [\href{https://arxiv.org/abs/astro-ph/9605054}{{\ttfamily astro-ph/9605054}}].

\bibitem{fixen_mather_2002_cobe}
D.J.~{Fixsen} and J.C.~{Mather}, \emph{{The Spectral Results of the Far-Infrared Absolute Spectrophotometer Instrument on COBE}}, \href{https://doi.org/10.1086/344402}{\emph{apj} {\bfseries 581} (2002) 817}.

\bibitem{Biyajima_2012}
M.~Biyajima and T.~Mizoguchi, \emph{{Analysis of residual spectra and the monopole spectrum for 3 K blackbody radiation by means of non-extensive thermostatistics}}, \href{https://doi.org/10.1016/j.physleta.2012.10.036}{\emph{Phys. Lett. A} {\bfseries 376} (2012) 3567} [\href{https://arxiv.org/abs/1210.5900}{{\ttfamily 1210.5900}}].

\bibitem{groeneveld_skewness_1984}
R.A.~Groeneveld and G.~Meeden, \emph{Measuring skewness and kurtosis}, {\emph{Journal of the Royal Statistical Society Series D: The Statistician} {\bfseries 33} (1984) 391}.

\bibitem{DeCarlo_kurtosis_1997}
L.T.~DeCarlo, \emph{On the meaning and use of kurtosis.}, {\emph{Psychological Methods} {\bfseries 2} (1997) 292}.

\bibitem{Dhal_calculation_2023}
S.~{Dhal}, S.~{Singh}, K.~{Konar} and R.K.~{Paul}, \emph{{Calculation of Cosmic microwave background radiation parameters using COBE/FIRAS dataset}}, \href{https://doi.org/10.1007/s10686-023-09904-w}{\emph{Experimental Astronomy} {\bfseries 56} (2023) 715}.

\bibitem{Dhal_dipole_2023}
S.~Dhal and R.K.~Paul, \emph{{Investigation on CMB monopole and dipole using blackbody radiation inversion}}, \href{https://doi.org/10.1038/s41598-023-30414-4}{\emph{Sci. Rep.} {\bfseries 13} (2023) 3316}.

\bibitem{hu_thermalisation_1993}
W.~Hu and J.~Silk, \emph{Thermalization constraints and spectral distortions for massive unstable relic particles}, \href{https://doi.org/10.1103/PhysRevLett.70.2661}{\emph{Phys. Rev. Lett.} {\bfseries 70} (1993) 2661}.

\bibitem{shaw_primordial_magnetism_2012}
J.R.~Shaw and A.~Lewis, \emph{Constraining primordial magnetism}, \href{https://doi.org/10.1103/PhysRevD.86.043510}{\emph{Phys. Rev. D} {\bfseries 86} (2012) 043510}.

\bibitem{ade_simons_obs_2019}
P.~Ade, J.~Aguirre, Z.~Ahmed, S.~Aiola, A.~Ali, D.~Alonso et~al., \emph{The simons observatory: science goals and forecasts}, {\emph{Journal of Cosmology and Astroparticle Physics} {\bfseries 2019} (2019) 056}.

\bibitem{paoletti_LiteBIRD_2022}
{\scshape LiteBIRD} collaboration, \emph{{The $LiteBIRD$ mission}}, \href{https://doi.org/10.22323/1.414.0085}{\emph{PoS} {\bfseries ICHEP2022} (2022) 085}.

\bibitem{khatri_constraint_2015}
R.~{Khatri} and R.~{Sunyaev}, \emph{{Constraints on {\ensuremath{\mu}}-distortion fluctuations and primordial non-Gaussianity from Planck data}}, \href{https://doi.org/10.1088/1475-7516/2015/09/026}{\emph{jcap} {\bfseries 2015} (2015) 026} [\href{https://arxiv.org/abs/1507.05615}{{\ttfamily 1507.05615}}].

\end{thebibliography}\endgroup



\end{document}